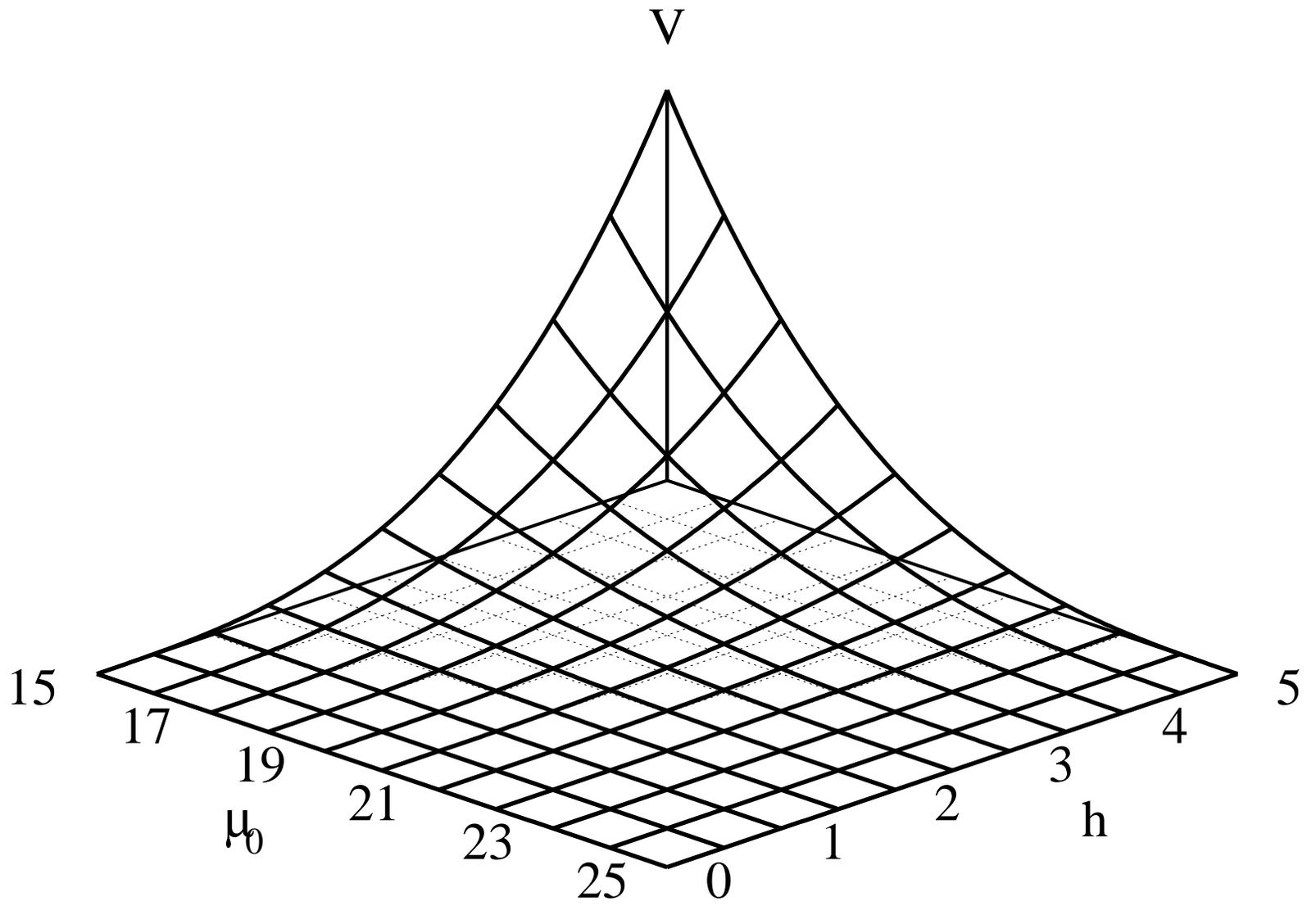

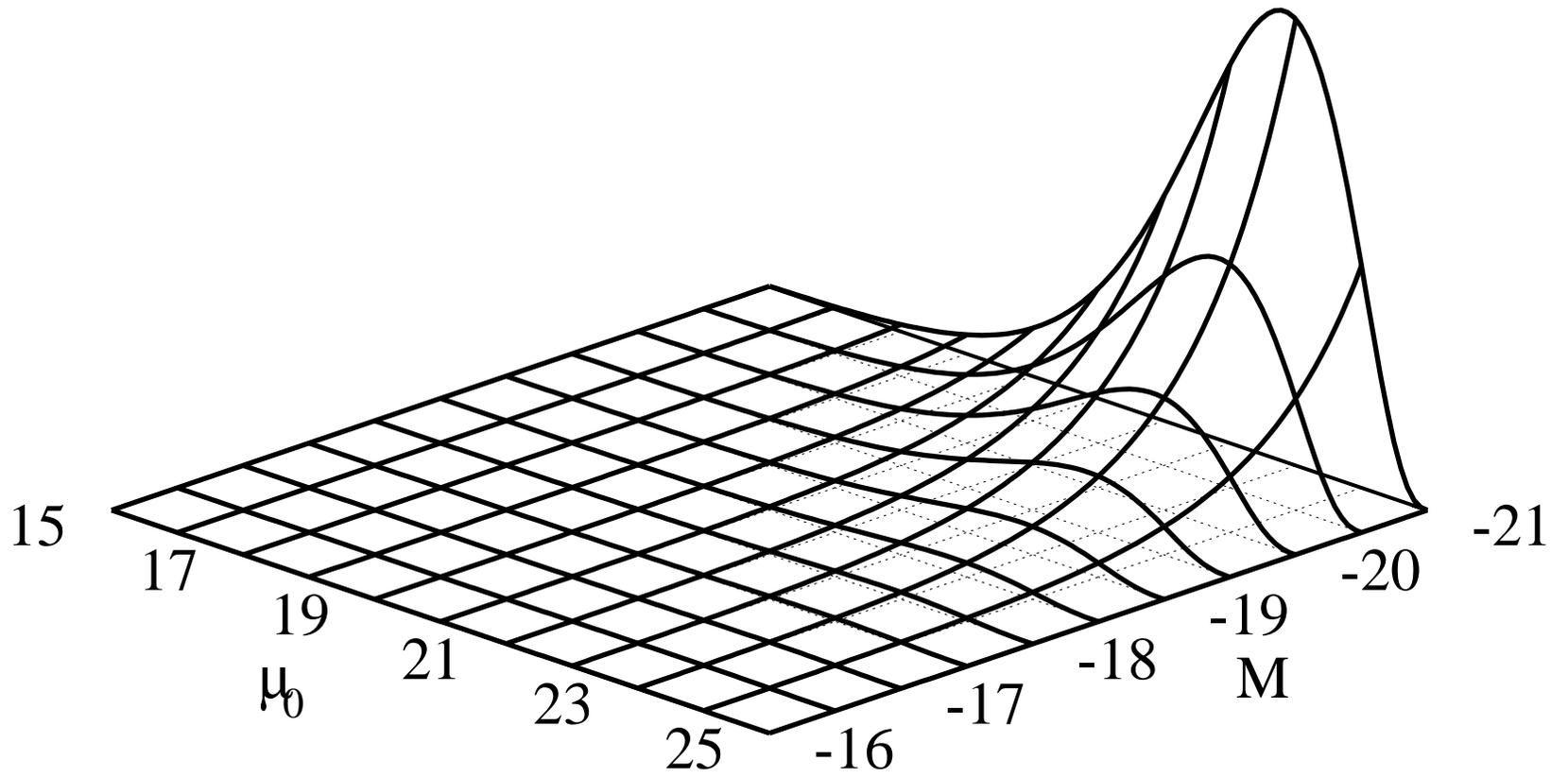

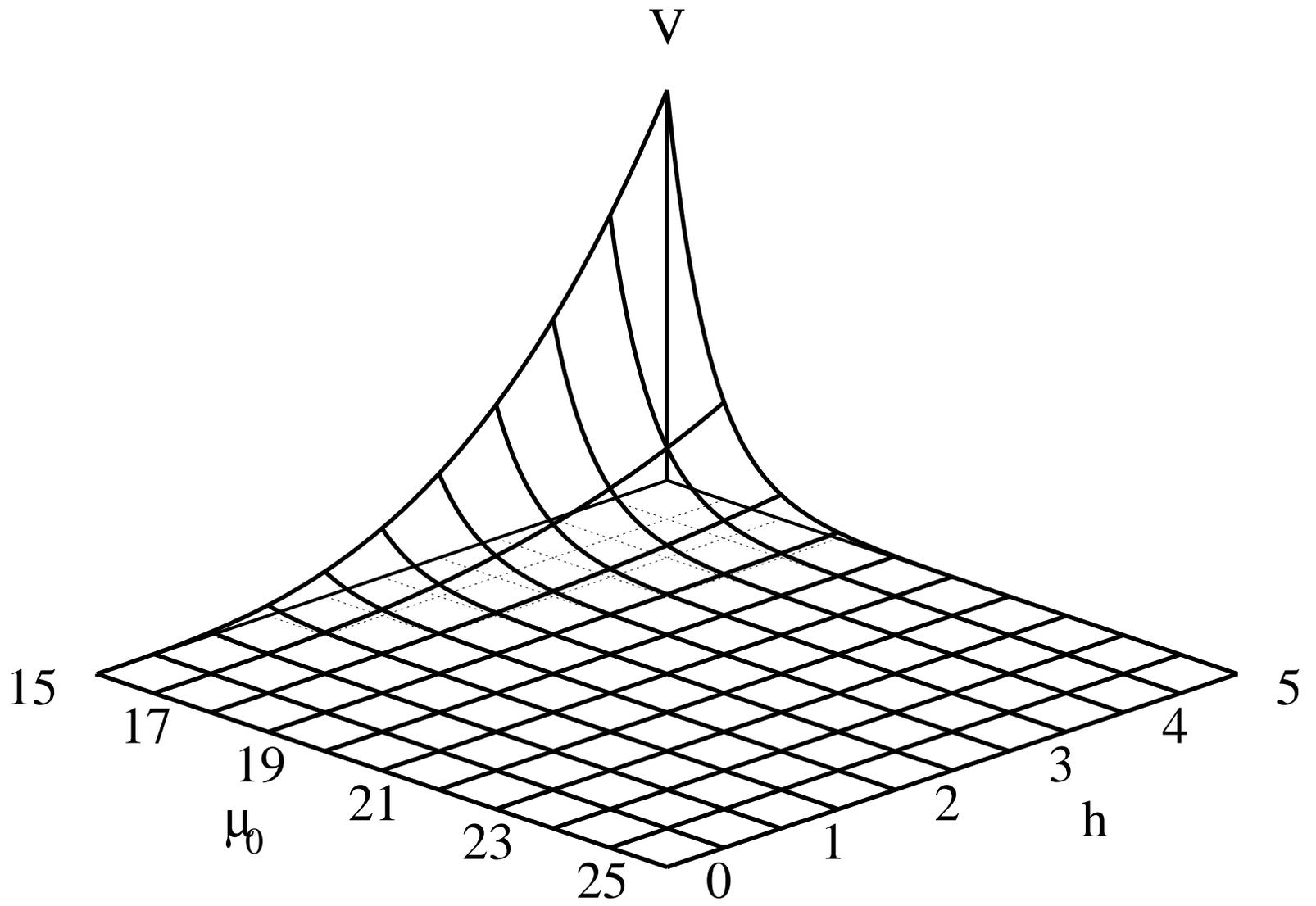

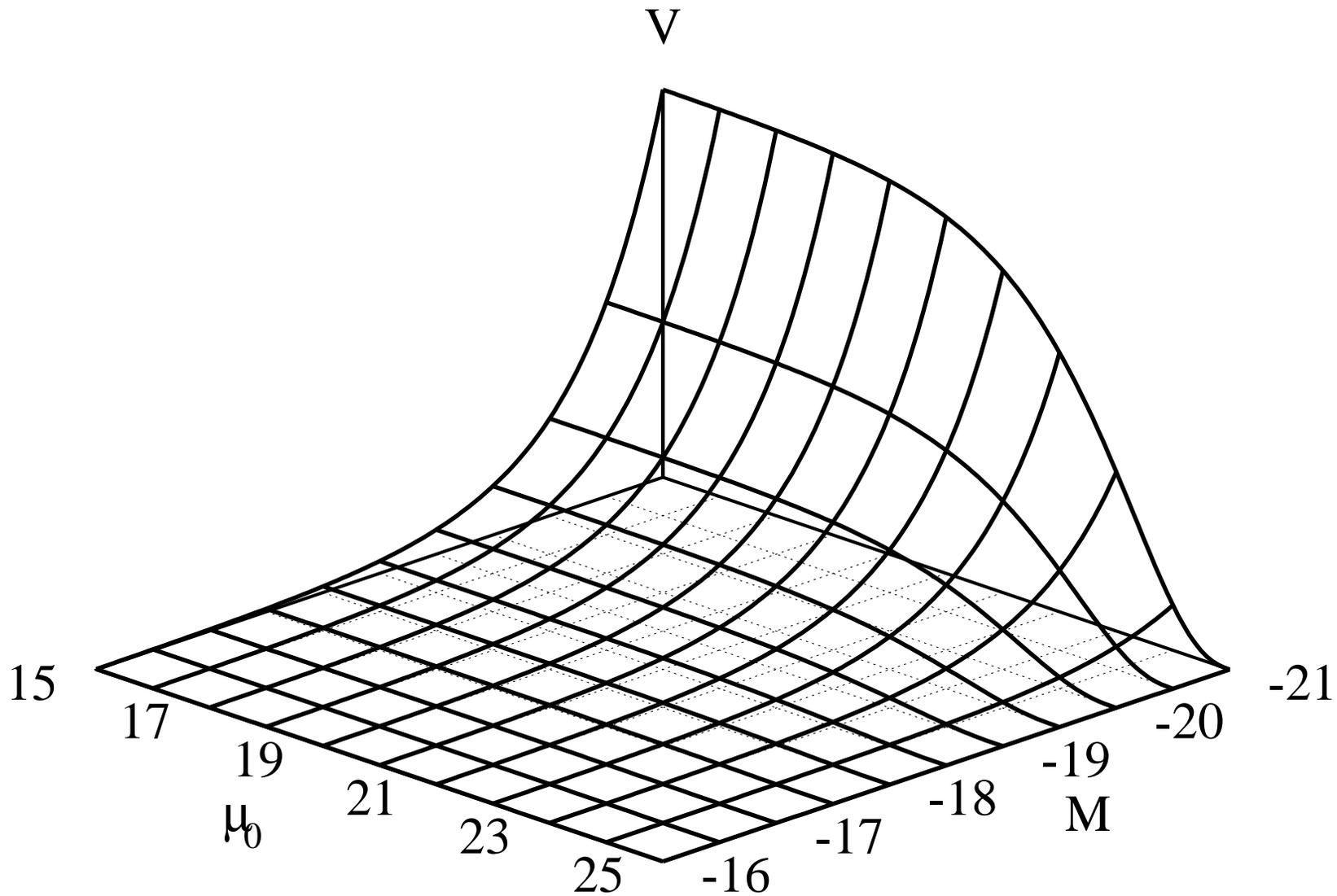

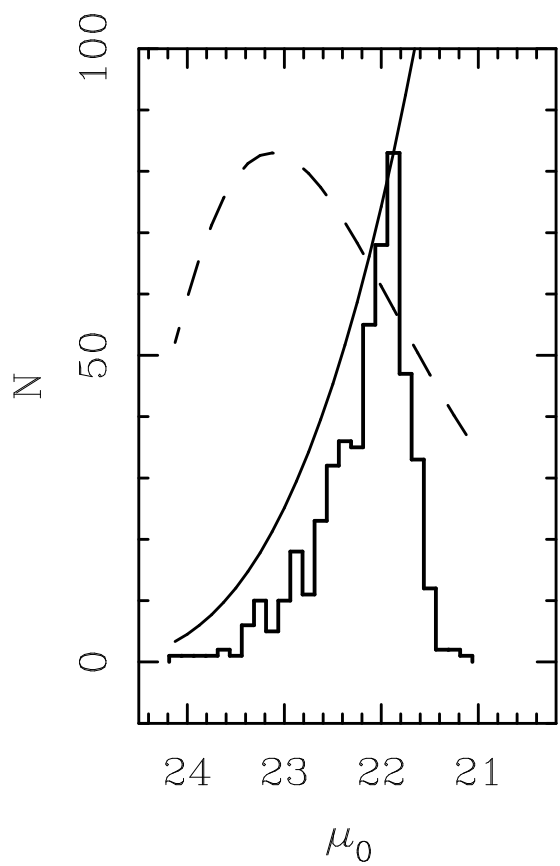
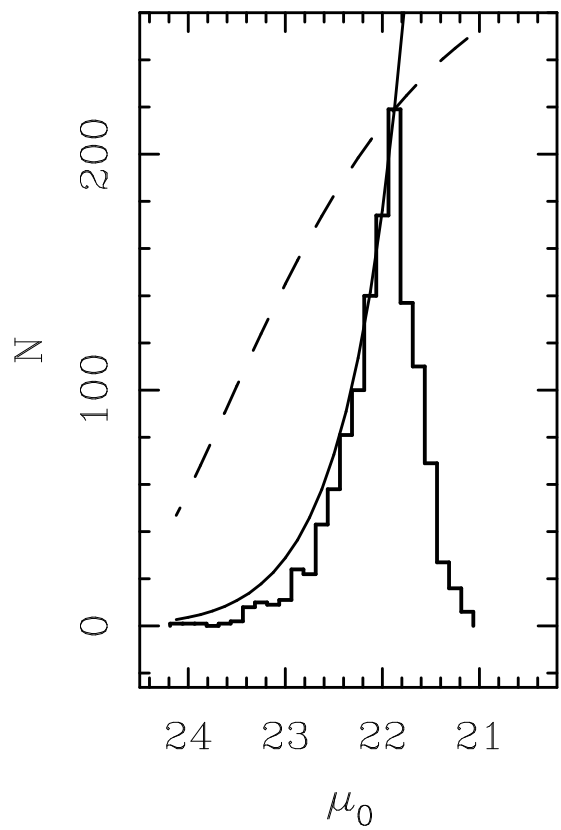
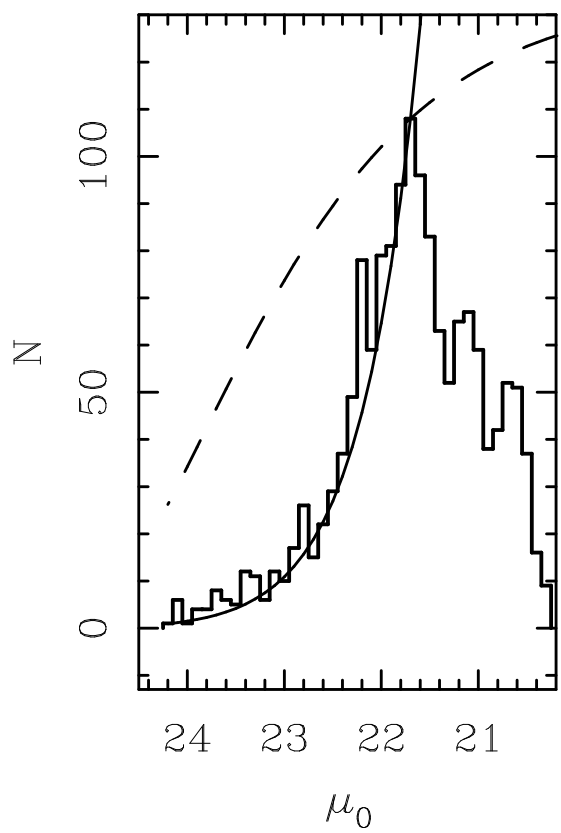
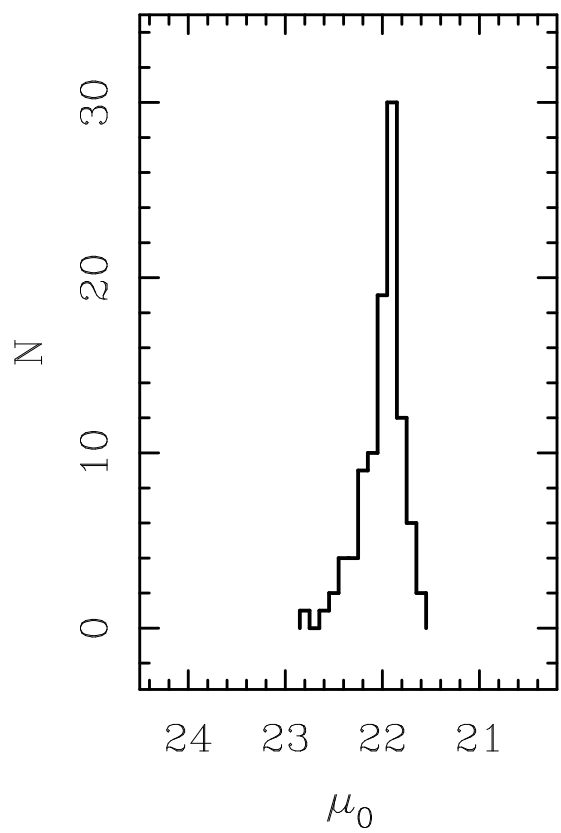

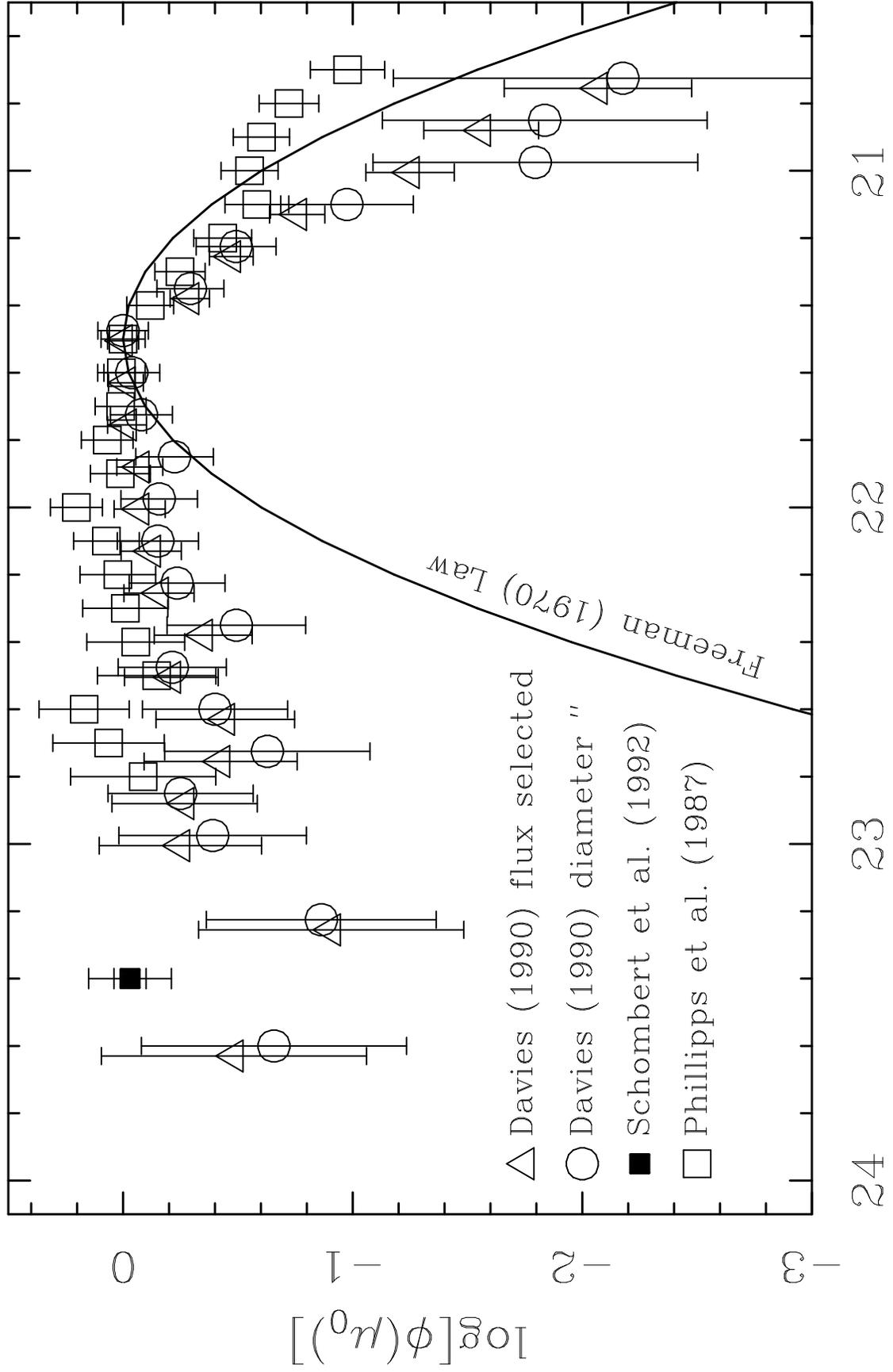

# Galaxy Selection and the Surface Brightness Distribution

Stacy S. McGaugh, [1] Gregory D. Bothun, [2] and James M. Schombert [3,4]

## ABSTRACT

Optical surveys for galaxies are biased against the inclusion of low surface brightness (LSB) galaxies. Disney (1976) suggested that the constancy of disk central surface brightness noticed by Freeman (1970) was not a physical result, but instead was an artifact of sample selection. Since LSB galaxies do exist, the pertinent and still controversial issue is if these newly discovered galaxies constitute a significant percentage of the general galaxy population. In this paper, we address this issue by determining the space density of galaxies as a function of disk central surface brightness. Using the physically reasonable assumption (which is motivated by the data) that central surface brightness is independent of disk scale length, we arrive at a distribution which is roughly flat (i.e., approximately equal numbers of galaxies at each surface brightness) faintwards of the Freeman (1970) value. Brightwards of this, we find a sharp decline in the distribution which is analogous to the turn down in the luminosity function at $L^*$. An intrinsically sharply peaked "Freeman Law" distribution can be completely ruled out, and no Gaussian distribution can fit the data. Low surface brightness galaxies (those with central surface brightnesses fainter than $22\ B\ \mathrm{mag\ arcsec}^{-2}$) comprise $\gtrsim 1/2$ the general galaxy population, so a representative sample of galaxies at $z = 0$ does not really exist at present since past surveys have been insensitive to this component of the general galaxy population.

[1]Institute of Astronomy, University of Cambridge, Madingley Road, Cambridge CB3 0HA, UK

[2]Physics Department, University of Oregon, Eugene, OR 97403, USA

[3]Infrared Processing and Analysis Center, California Institute of Technology, Pasadena, California 91125, USA

[4]Present address: Astrophysics Division, Code SZ, NASA HQ, Washington, D.C., 20546

## 1. Introduction

Galaxies, and nebulae in general, are by observational definition extended, resolved objects seen in projection against a noisy background of finite brightness. This fundamental difference from stellar point sources results in a difference in the visibility of nebulae. For stars, the only quantity relevant to selection is the total flux within a survey's passband. For extended objects, not only must the total flux be considered, but also the way in which that flux is distributed across the object, and the percentage of the total flux which is above some isophotal background level.

In practice, this can be rather complicated. The simplest interesting case beyond a point source is that of an object with an azimuthally symmetric radial light profile. Though galaxies are not completely described by such profiles, they are reasonably approximated by them. The assumption of azimuthal symmetry reduces the number of parameters relevant to selection to three: the characteristic size of the object, the characteristic surface brightness, and the shape of the profile.

For galaxies, the profile shape is usually assumed to be either exponential (for disks) or $r^{1/4}$ (for ellipticals). Within this idealized framework of azimuthally symmetric galaxy profiles, we shall focus on disk dominated systems, thus reducing the number of selection parameters to two. The exponential profile, in magnitude units, is

$$\mu(r) = \mu_0 + 1.086 \frac{r}{\alpha}, \qquad (1)$$

where $\mu_0$ is the central surface brightness of the disk and $\alpha$ is its angular scale length which corresponds to the physical scale length $h$ at distance $d$. These two parameters characterize the light distribution of the idealized disk galaxy, and together determine the integrated luminosity,

$$L = 2\pi h^2 \Sigma_0 f(x). \qquad (2)$$

Here $\Sigma_0$ is the central surface brightness in linear units, and

$$f(x) = 1 - (1+x)e^{-x} \qquad (3)$$

gives the fraction of the light contained within a finite number of scale lengths $x$,

$$x = \frac{r}{\alpha} = 0.92[\mu(r) - \mu_0] \qquad (4)$$

relative to that contained in an exponential profile extrapolated to infinity (Disney & Phillipps 1983, hereafter DP). These simple formulas provide adequate fits to most spiral galaxies (de Vaucoulers 1959), and in particular to low surface brightness (LSB) galaxies (i.e., those with $\mu_0 > 23$: McGaugh & Bothun 1994; Sprayberry et al. 1995; de Blok et al. 1995).

## 2. Volume Sampling and Central Surface Brightnesses

Freeman (1970) found that all spiral disks have essentially the same central surface brightness, $\mu_0 = 21.65 \pm 0.3$ $B$ mag arcsec$^{-2}$. This has become known as "Freeman's Law." If correct, the number of parameters relevant to galaxy selection reduces to one as only variations in size modulate those in luminosity.

Since $\mu_0$ is a measure of the characteristic surface mass density of a disk, Freeman's Law places very stringent requirements on the physical processes of galaxy formation and evolution in order to result in this specific value for all spirals. Variations in the mass to light ratio, star formation history, collapse epoch, and initial angular momentum content must all conspire to balance at this arbitrary value. It is thus important to rigorously test the reality of Freeman's Law as the distribution of $\mu_0$ may be directly related to the conditions of galaxy formation (Freeman 1970; van der Kruit 1987; McGaugh 1992; Mo *et al.* 1994).

The Freeman value is about 1 magnitude brighter than the surface brightness of the darkest night sky. That the number of galaxies with faint central surface brightnesses appears to decline rapidly as $\mu_0 \to \mu_{sky}$ is suspicious and if true of the real galaxy population implies that our observational viewpoint is privileged in that we are capable of detecting most of the galaxies that exist, at least when the moon is down. These concerns were raised by Disney (1976), who suggested that the Freeman (1970) result could stem from selection effects. This line of reasoning was formalized by DP, who quantified the volume sampled by galaxy surveys as a function of central surface brightness (which they call the "visibility"). This is the sampling function which, after convolution with the intrinsic galaxy distribution, yields the apparent distributions found in specific catalogs. However, Allen & Shu (1979) pointed out that the selection effects involved were unlikely to behave in precisely the manner described by Disney (1976), which is the basis of the DP formalism. Here, we rederive the visibility as a function of the two disk parameters and make an explicit comparison to the formalism of DP.

### 2.1. Diameter Limit

The two selection parameters which need to be specified for a galaxy catalog selected by diameter are the diameter limit $\theta_\ell$ and the isophotal level $\mu_\ell$ at which the diameter is measured. The requirement is that $\theta = 2r \geq \theta_\ell$ when $\mu(r) = \mu_\ell$. From equation (1) it follows that

$$\theta = 1.84\alpha(\mu_\ell - \mu_0) \propto \frac{h}{d}(\mu_\ell - \mu_0). \tag{5}$$

The maximum distance at which a galaxy can lie and meet the selection criteria occurs when $\theta = \theta_\ell$, so

$$d_{max} \propto \frac{h}{\theta_\ell}(\mu_\ell - \mu_0). \tag{6}$$

The volume sampled as a function of the parameters that describe a galaxy is

$$V(h, \mu_0) \propto d_{max}^3 \propto h^3(\mu_\ell - \mu_0)^3. \tag{7}$$

This is ploted in Fig. 1(a) for the case $\mu_\ell = 25$ mag arcsec$^{-2}$. The precise value of $\mu_\ell$ is irrelevant, but this is typical of surveys for bright galaxies.

In contrast, DP derive the visibility as a function of surface brightness at fixed luminosity, then state that this can be scaled by luminosity (i.e., $V \propto L^{3/2}$). For a diameter limited catalog, they find that the volume sampled goes as

$$V(\mu_0) \propto (\mu_\ell - \mu_0)^3 10^{-0.6(\mu_\ell - \mu_0)} \tag{8}$$

at fixed luminosity (see their equation 31). In order to show the full functional dependence on the two parameters which they use to describe the disk, we scale by luminosity as they suggest, arbitrarily normalized at $M^* = -21$ as they chose to do. Their full visibility, including the effects of both central surface brightness and absolute magnitude, is thus

$$V(M, \mu_0) \propto (\mu_\ell - \mu_0)^3 \, 10^{-0.6[(\mu_\ell - \mu_0) + (M - M^*)]}. \tag{9}$$

This is shown in Fig. 1(b).

Equation (9) is mathematically equivalent to equation (7) upon transformation of variables between $M$ and $h$. However, Figures 1(a) and (b) give a very different impression of how galaxy selection works. The important difference between Figures 1(a) and (b) is conceptual rather than mathematical. The question we wish specifically to answer is *how does the volume sampled by a survey depend on the intrinsic properties of any particular object?* For point sources, this is a one parameter problem completely specified by the luminosity of an object. For extended sources like galaxies, the total luminosity is not so simply related to the observable quantities and must be decomposed into the component parameters which describe the object (e.g., equations 1 – 4). The correct physical basis should not, therefore, include the luminosity itself, but rather these component parts. To appreciate the conceptual distinction, imagine changing the surface brightness of a galaxy (or any extended object of characteristic size $h$). As $\mu_0$ varies, $M$ varies with it but $h$ remains fixed. In order to hold $M$ fixed, as DP did, we must artificially adjust the size of the object so that it is no longer physically the same object. Hence equation (9) does not address quite the correct question, and as a result the axes of Fig. 1(b) are not composed of orthogonal, independent quantities.

The DP visibility function thus gives a very misleading impression which has led to a number of misconceptions about this issue. Most notably, the broad peak in Fig 1(b) always occurs at $\mu_0 = \mu_\ell - 2.17$. That the value of $\mu_0$ favored in this way is very near the Freeman value if $\mu_\ell \approx 24$ is the root of Disney's argument. Thus, a robust prediction of the DP visibility formalism is that the central surface brightness typically found in diameter limited surveys will grow fainter as surveys are pushed deeper. Contrary to this expectation, the peak in the apparent distribution is observed *not* to vary with $\mu_\ell$ (Phillipps *et al.* 1987; van der Kruit 1987). Moreover, the observed distribution is too narrow to be explained by the broad peak (as noted by DP.)

There is no peak at a preferred surface brightness in Fig. 1(a), with the volume probed increasing without bound as the surface brightness becomes brighter. The variation of $V$ with $\mu_0$ is extremely rapid, so we *expect* that the apparent surface brightness distribution should always be very strongly peaked around the brightest value which exists in the intrinsic distribution, *regardless* of the value of $\mu_\ell$, as observed. Contrary to the argument of Disney (1976), very high surface brightness galaxies would be easily detected if they existed (cf. Allen & Shu 1979). Very *small* galaxies might be missed, but this is a separate issue. Size and surface brightness form the most natural orthogonal basis for purposes of selection and there is much empirical evidence to support this (see below). Thus, it is necessary to fully consider their separate effects on galaxy surveys.

### 2.2. Flux Limit

For flux limited samples where isophotal magnitudes are employed, the selection parameters are the magnitude limit $m_\ell$ and the isophotal level $\mu_\ell$ above which the flux is measured. Catalogs of galaxies limited by total flux do not exist, since survey material always has an effective isophotal limit below which very diffuse galaxies can not be identified, regardless of their total flux. In fact, these galaxies do exist (e.g., Malin 1 and its cousins; Bothun *et al.* 1987; Impey & Bothun 1989); Malin 2 (Bothun *et al.* 1990) has an apparent magnitude of $B = 14.2$ but is contained in neither the NGC, which contains many fainter galaxies, nor the UGC (Nilson 1973), which in addition to the limit $\theta_\ell = 1'$ is also supposedly complete to the usually less demanding limit $B_\ell = 14.5$.

The volume sampled depends on the portion of the luminosity visible above the isophotal level of selection. That is, $V \propto L_\ell^{3/2}$, where $L_\ell$ can be decomposed into $\mu_0$ and $h$ using equation (2):

$$V \propto [L(r < \theta_\ell/2)]^{3/2} \propto [\Sigma_0 h^2 f(x_\ell)]^{3/2} \propto h^3 10^{-0.6(\mu_0 - \mu_0^*)} [f(x_\ell)]^{3/2} \qquad (10)$$

where $\mu_0^*$ is an arbitrary normalization factor. For convenience we will take it to be the bright end cutoff value in the intrinsic distribution, in analogy with $L^*$. Again, $\mu_0$ and $h$ form the natural orthogonal bases yielding the form shown in Fig. 1(c).

For flux selection, DP (again, see their equation 31) give

$$V(M, \mu_0) \propto 10^{-0.6(M - M^*)} [f(x_\ell)]^{3/2}, \qquad (11)$$

which is plotted in Fig. 1(d). Though mathematically equivalent, retaining the luminosity as the variable of interest rather than its component parts again gives a misleading result. One curious example of this is that equation (11) is often erroneously stated as $V \propto L^{3/2} \Sigma_0^{3/2}$. In effect, DP only consider the fraction of the total flux $f(x_\ell)$ which is detectable above the isophotal level of selection (a relatively slowly varying function of $\mu_\ell$), and not the fact that the luminosity of a galaxy decreases with its surface brightness. Again, computing the variation of the visibility with surface brightness *at fixed luminosity* is conceptually incorrect.

Note that flux selected samples will have apparent distributions of $\mu_0$ which are even more strongly peaked around the brightest extant value $\mu_0^*$ than diameter limited samples, because the factor $10^{-0.6(\mu_0-\mu_0^*)}$ varies more rapidly than does $(\mu_\ell - \mu_0)^3$. Magnitude limited samples will detect more galaxies in total than diameter limited ones at any given value of $\mu_\ell$, simply because they admit very distant, intrinsically luminous galaxies. They will always be strongly dominated by the largest ($h^*$), highest surface brightness ($\mu_0^*$), and hence $L^*$, galaxies. Diameter selection yields samples which are less biased and more representative of the general field population (witness the significant number of LSB galaxies contained in the UGC: Romanishin *et al.* 1983; McGaugh & Bothun 1994; de Blok *et al.* 1995).

In either case, the volume over which low surface brightness galaxies can be detected is very small. It goes to zero if the central surface brightness happens to be fainter than the selection isophote, even if the galaxy in question is intrinsically luminous. Examples of luminous galaxies with such faint central surface brightnesses are known to exist. Hence, given the relatively bright effective selection isophotes of large area surveys, complete flux limited samples of galaxies by definition do not exist. Moreover, it is necessary to give at least a two dimensional description of the completeness limit of a survey, not just a magnitude limit. Equivalently, as stressed by Ellis & Perry (1979), at least an area as well as a flux should be reported by surveys, as not to do so loses information. These effects are already a problem in local surveys, and become particularly severe in the cosmological context (Ellis *et al.* 1984; Sievers *et al.* 1985; McGaugh 1994).

## 3. The Surface Brightness Distribution

Since the volume over which low surface brightness galaxies can be observed in any given survey is so small, the very existence of such objects requires a large space density. The volume sampled for disk galaxies with $\mu_0 = 21.5$ is always very much larger than that sampled for galaxies with $\mu_0 = 23.5$. Obviously this biases all surveys towards incorrectly concluding that the space density of LSB galaxies is small. No consideration of these effects were made by Freeman (1970) or in some later treatments (e.g., Bosma & Freeman 1993), while in others (as described above) they are seriously mistaken. A correction for volume sampling effects *must* be applied to any survey; without it we would, for example, conclude that K-giants are the most common type of star in the Galaxy.

Ideally, we should determine the bivariate galaxy distribution $\Phi(h, \mu_0)$ from complete catalogs for which the selection parameters $\mu_\ell$ and $\theta_\ell$ or $m_\ell$ are carefully specified and rigorously applied. Note that it is not possible to eliminate $\mu_\ell$ by selecting by both diameter and magnitude, since this begs the question of what precisely is being measured for each. Large catalogs which obey these strict criteria do not yet exist. Another requirement for the measurement of the bivariate distribution is that surface photometry be performed on all objects (i.e., $\mu_0$ and $h$ must actually be measured). This has not often been done, and satisfying these criteria is really the best reason for performing a uniform digital sky survey.

A final requirement for measuring $\Phi(h,\mu_0)$ is that redshifts be measured in order to extract absolute information (i.e., $h$ instead of $\alpha$). No samples exist which meet all these requirements, though some come close (de Jong & van der Kruit 1994). Phillipps *et al.* (1987) and Davies (1990) do present data which meet the requirements for rigorous selection and measurement of $\mu_0$, lacking only redshifts. These data sets consist of complete samples of several hundred galaxies selected by isophotal magnitude in the case of Phillipps *et al.* (1987) and both isophotal magnitude and diameter in the case of Davies (1990). The survey of Phillipps *et al.* (1987) is in the direction of the Fornax cluster, but all higher surface brightness galaxies (those with $\mu_0 < 23$) are expected to be in the background field (Ferguson 1989; Irwin *et al.* 1990). Davies (1990) surveyed both Fornax and the adjacent field; we are concerned only with the field data. The isophotal level of selection of Phillipps *et al.* (1987) is $\mu_\ell = 25.5$ mag arcsec$^{-2}$ and that of the Davies (1990) field data is $\mu_\ell = 25.3$, both in the $B_J$ band.

The relevant input data are the number of galaxies detected at each central surface brightness, $N(\mu_0)$ (i.e., the apparent distribution of $\mu_0$). These are shown in Fig. 2, together with the expectation from the DP formalism and that derived here. The DP visibility function fails to predict the shape of the apparent distributions. However, the data are well quite well described by the volume sampling function derived here. Surface brightness selection dominates the shape of the faint end of the apparent distributions, while there is a real maximum to the surface brightness distribution (as first noted by Allen & Shu 1979).

The data set which currently has the most constraining power is provided by Schombert *et al.* (1992). This catalog contains a large number ($\sim 200$) of LSB galaxies (most with known redshifts) with $\theta \geq 1'$ measured at $\mu_\ell = 26$ $B$ mag arcsec$^{-2}$. This catalog is characterized by disk galaxies of typical ($h^*$) size but low surface brightness, having a distribution very sharply peaked at $\mu_0 = 23.4$ (McGaugh & Bothun 1994; de Blok *et al.* 1995). As noted by Schombert *et al.* (1992), the very existence of so many of these galaxies $> 4\sigma$ from the Freeman value is inconsistent with a Freeman Law. Indeed, the major uncertainty in estimating their space density comes in finding a comparison sample of galaxies which are actually known to *obey* Freeman's Law (see McGaugh 1995).

Since surface brightness is distance independent, we do not need redshifts to derive the surface brightness projection of the bivariate distribution. The relative number density of disk galaxies as a function of central surface brightness,

$$\phi(\mu_0) = \frac{N(\mu_0)}{N(\mu_0^*)} \frac{V(h^*,\mu_0^*)}{V(h,\mu_0)}, \qquad (12)$$

follows from the apparent distribution corrected for volume sampling. Since no absolute information is available without redshifts, we determine the relative distribution by normalizing to an arbitrary fiducial galaxy of parameters $(h^*, \mu_0^*)$. Since only relative volumes normalized to these fiducial values are involved, the resulting volume correction is very accurate and not subject to the statistical estimation problems encountered when constructing absolute quantities like the luminosity function.

The surface brightness distribution follows directly from the observations $[N(\mu_0)$ an $\mu_\ell]$ and equation (12) given one assumption. The volume correction factor depends on $h$ as well as $\mu_0$, so it is necessary to make an assumption about $h$. Though it would obviously be preferable to determine the full bivariate distribution, the necessary data do not exist. So, in order to make progress, we assume that scale length is not correlated with central surface brightness. Thus, at any $\mu_0$, the effects of volume sampling due to variations in $h$ on average cancel out, and only those due directly to $\mu_0$ matter. Hence we make the approximation

$$\frac{V(h^*, \mu_0^*)}{V(h, \mu_0)} \approx \frac{V(\mu_0^*)}{V(\mu_0)}. \tag{13}$$

This is the most natural assumption to make given the form of equations (7) and (10). More importantly, the assumption that scale length is uncorrelated with central surface brightness is borne out by a wealth of observational data (Romanishin *et al.* 1983; Davies *et al.* 1988; Irwin *et al.* 1990; McGaugh & Bothun 1994; Sprayberry *et al.* 1995; de Blok *et al.* 1995; de Jong 1995; McGaugh *et al.* 1995). Even if this assumption were incorrect, it does not alter the basic conclusion that there must be a relatively large space density of LSB galaxies, simply because $V^{-1} \to \infty$ as $\mu_0 \to \mu_\ell$. The detection of any galaxy with a central surface brightness approaching the selection isophote immediately implies a large density of such objects.

If there *is* a correlation between $\mu_0$ and $h$ in the sense that galaxies with faint central surface brightnesses are on average smaller, then there is an additional factor working against their selection, thus requiring even more of them. If, on the other hand, the trend is in the opposite sense with $h$ on average being larger as $\mu_0$ becomes fainter, then there are not quite as many LSB galaxies. However, they still must be much more numerous than suggested by a Freeman Law, and each one is large and intrinsically luminous in the same sense as Malin 1 (if not necessarily as extreme).

Also implicit in equation (12) is the assumption that a fair volume of space has been surveyed. Variations due to large scale structure may cause marginal variations in the distributions derived from the different data sets, but are very unlikely to be of the amplitude required to offset the large volume corrections required for LSB galaxies. It is even less likely that this could conspire to cause a false enhancement of their numbers, particularly for the survey of Schombert *et al.* (1992) who find a high surface density of LSB galaxies over a large area ($> 2000$ square degrees of sky).

Cosmological dimming might be invoked to explain the faintest points in the Phillipps *et al.* (1987) and Davies (1990) data, since these are selected at $B_J < 19$ where the median redshift is $z = 0.1$. However, this effect works both ways. Since LSB galaxies can only be discovered over small volumes, they will typically have low redshifts (see also McGaugh 1995). The distribution expcted from the dimming of 100 pure Freeman disks for the empirical redshift range (Koo & Kron 1992) appropriate to these data is shown in Fig. 2(d). This does not reproduce the shape of the actual data, which *is* well predicted by the *expected* $V(\mu_0)$. In any case, dimming can not explain the large number of bona-fide local, fairly luminous LSB galaxies catalogued by Schombert *et al.* (1992).

Additional support for our approach comes from the fact that it is consistent with extant determinations of the bivariate distribution (Sodré & Lahav 1993; de Jong 1995). The resultant surface brightness distribution is shown in Fig. 3. The data have a long, roughly flat tail towards lower surface brightness. That is, approximately equal numbers of disk galaxies exist at each central surface brightness. This is only true faintwards of the Freeman value, which we have chosen as the fiducial $\mu_0^*$. Brighter than this, there is a sharp cutoff. Though the extant data are not in perfect agreement as to how steep this cutoff is, there is a clear turndown. As expressed by Allen & Shu (1979), this is the proper physical interpretation of Freeman's Law. It is really a statement analogous to the fact that flux selected catalogs are always dominated by the brightest objects which exist in the intrinsic distribution (i.e., $L^*$ galaxies). Just as galaxies brighter than $L^*$ are rare, so too are galaxies with surface brightnesses higher than $\mu_0^*$. Diffuse galaxies do exist in substantial numbers, they are just harder to see.

## 4. Discussion

The Gaussian surface brightness distribution advocated by Freeman (1970) fails seriously to describe the true intrinsic distribution (Fig. 3). It underestimates the number of galaxies with $\mu_0 > 23$ mag arcsec$^{-2}$ by over five orders of magnitude. No adjustment to the assumption about the scale length distribution made here can reconcile the data with a Freeman Law. Increasing the dispersion of a Gaussian distribution (van der Kruit 1987) misses the basic point that the distribution is clearly not symmetric. The monumental difference between what Freeman's Law predicts and what is actually observed is a strong testimony to the fact that a proper survey for galaxies has yet to be done. The constancy of disk galaxy number density as a function of $\mu_0$ is the most significant result obtained to date on LSB galaxies and strongly alters the conventional view of the galaxy population, which suggests that they are at most a few percent of galaxies by number. In light of the evidence shown in Fig. 3, we find it difficult to believe that LSB galaxies can continue to be regarded as an unimportant constituent of the universe.

The realization that the surface brightness distribution is not just a Gaussian about some preferred value is analogous to the same realization for the luminosity function (Zwicky 1957; Schechter 1976), and has some very important consequences. The constancy of $\mu_0$ is a fundamental assumption in many important problems in extragalactic astronomy. For example, it is implicit in determinations of the field galaxy luminosity function, which is very sensitive to the failure of this assumption (McGaugh 1994; Ferguson & McGaugh 1995). Any errors introduced into the luminosity function propagate into calculations involving it, such as the number counts of galaxies and the cross section of galaxies as Ly$\alpha$ absorbers along sight lines to QSOs. Properly, these quantities can only be obtained by integrating over the bivariate distribution. (For Ly$\alpha$ absorbers really the bivariate distribution of the *gas* disks is required.)

The reason for this is simple. When performing these calculations, it is always necessary to make some approximation at the faint end of the luminosity function (either by truncation or extrapolation) to account for the faintest objects which are unconstrained by observations. In principle, the same must be done for the surface brightness portion of the bivariate distribution. The number of objects with $\mu_0 > 23.5$ mag arcsec$^{-2}$ is unknown, and can only be guessed by extrapolation of the trend in Fig. 3. A few examples of these objects have turned up in photographically based surveys to date, but such surveys are not very sensitive to the presence of this population. Significant numbers of galaxies with $\mu_0 > 24.0$ mag arcsec$^{-2}$ have been detected in recent CCD surveys, suggesting that the trend remains fairly constant (O'Neil *et al.* 1995) or even rises towards fainter $\mu_0$ (Schwartzenberg *et al.* 1995). There is no automatic requirement that these very LSB galaxies are low luminosity or gas poor, so it is possible that they make a significant contribution to these sorts of calculations. Currently, the extrapolation in the surface brightness dimension is not explicitly made as with the faint end of the luminosity function; rather, very LSB galaxies are implicitly assumed not to exist. When surface brightness selection effects are properly accounted for, the change in the faint end slope of the luminosity function can be dramatic (see Bothun *et al.* 1991).

## 5. Conclusions

We have demonstrated that disk galaxies exist in approximately equal numbers at every surface brightness fainter than a maximum value $\mu_0^*$. This resolves the long standing controversy over the number density of low surface brightness galaxies. These do exist in substantial numbers as suspected by Disney (1976), but the selection effects involved do not work in precisely the way he described. This led to a misleading, demonstrably incorrect (van der Kruit 1987) prediction for what should be observed. The correct expectation (cf. Allen & Shu 1979) shows that since Freeman's (1970) Law refers only to the apparent distribution, it is indeed a selection effect. Applying an appropriate volume correction directly yields the true distribution.

Our result on the space density of galaxies as a function of central surface brightness requires a basic adjustment in the way we think about galaxies. Much of the thought in the field is implicitly one dimensional, with one parameter (like luminosity or morphology) dominating the way problems are approached. This is just not sufficient, as at least two parameters are required to convey a minimally useful amount of information about complicated objects like galaxies. Obviously, more are preferable, but a minimum of $\mu_0$ and $\alpha$ are required to describe galaxy selection. Perhaps with the large digital surveys currently in the pipeline it will be possible to rigorously apply strict selection criteria and develop methods to more fully characterize the appearance of galaxies and all the parameters relevant to their selection. At the very least, these surveys will help to establish a more representative sample of galaxies than has been obtained to date.

That the known examples of low surface brightness galaxies have many properties which are very different from what is considered 'normal' (for Freeman disks), and yet that they exist in approximately equal numbers, indicates that a great deal more has yet to be learned about the local galaxy population. Until the properties of local galaxies are better quantified, it is impossible to sensibly interpret observations at high redshift, much less use these to constrain evolution and cosmological models. Many outstanding problems, like those of the baryonic missing mass, faint blue galaxies, and Ly$\alpha$ absorbers, though not entirely solved by the substantial population of low surface brightness galaxies, are certainly mollified. Models which address these problems generally do not consider the possibility of a broad distribution of surface brightnesses. Rather, they implicitly assume the incorrect Freeman Law.

In order to adequately characterize the local galaxy population, it is necessary to perform a survey which

1. is complete to rigorously defined and applied limits,

2. explicitly quantifies a uniform isophotal level at which fluxes or diameters are measured, and

3. actually characterizes galaxy images with at least two parameters such as the central surface brightness and scale length.

More parameters are of course needed to fully describe the appearance of galaxies, but these two are a minimum requirement for the purposes of selection. A parameter describing the profile shape is probably required as well for flux limited surveys in order to adequately account for bulge components. This is less important for diameter limited surveys which only depend on the behavior of the profile at large radii. These can always be well approximated by a straight line fit through the neighboring isophotes. Diameter limited surveys are also superior in that they are less biased in favor of the brightest objects. Until such surveys are performed, the extragalactic community should not be surprised when new galaxies are found. Indeed, the recent discoveries of LSB galaxies have now given us a substantially different view of the general galaxy population than existed just a decade ago. With no sign of a turn down in number density at low surface brightness levels (e.g., Fig. 3), there are clearly more nearby galaxies awaiting discovery.

We are grateful to Roelof de Jong, Steve Phillipps, Harry Ferguson, and Greg Aldering for discussions related to the surface brightness problem. We also thank the referee for a number of positive suggestions.

---

This preprint was prepared with the AAS LaTeX macros v3.0.

Fig. 1.— Galaxy selection in terms of the volume probed as a function of the two parameters which are used to describe a galaxy. In (a) and (b), selection is by isophotal diameter. In (b) and (c) it is by isophotal magnitude. Panels (a) and (c) show equations (7) and (10), respectively. These give the volume sampled as a function of the orthogonal properties of size $h$ and central surface brightness $\mu_0$. From (a) and (c) it is clear that the volume sampled by a survey increases monotonically as size and surface brightness increase. Small and low surface brightness galaxies will thus be underrepresented in complete catalogs. The formalism of DP is shown in (b) and (d) (equations 9 and 11, respectively). The two parameters plotted here, absolute magnitude and surface brightness, are not orthogonal properties of a galaxy, so these plots give a misleading impression. For example, there is no selection effect acting against high surface brightness galaxies as implied by (b). Size and surface brightness are separate issues. In addition, (b) and (d) predict that observed distributions should have a broad peak, the position of which which (for b) varies with the isophotal selection level. In contrast, (a) and (c) predict that the apparent distribution will always have a narrow peak at the brightest central surface brightness which exists in the intrinsic distribution. This latter behavior is what is observed. Note also that diameter limited surveys are less biased than those limited by flux.

Fig. 2.— The number of galaxies observed at each central surface brightness. The field data of Davies (1990) are selected by (a) diameter and (b) flux. The data of Phillipps et al. (1987) are selected by flux (c) in the direction of the Fornax cluster, but galaxies with $\mu_0 > 23$ are in the background field. The smooth solid lines are not fits to the data, but simply what is expected from the volume sampling function derived here. The data are closely matched by these selection effects combined with a real cutoff in the intrinsic distribution at high surface brightnesses. The dashed lines show the predictions of the DP visibility formalism. The data are not well predicted by DP or by the cosmological dimming expected for pure Freeman disks (d) for the redshifts appropriate to the selection magnitude ($B_J < 19$) of panels (b) and (c).

Fig. 3.— The surface brightness distribution, giving the relative numbers of galaxies at each central surface brightness. Open symbols: the differential distribution obtained from the data in Fig. 2 corrected for volume sampling effects assuming that $\mu_0$ is not correlated with $h$. The data have been shifted along the abscissa to minimize overlap. Solid symbol: estimate of the density of low surface brightness galaxies cataloged by Schombert et al. (1992). Error bars are from counting statistics. For the Schombert et al. (1992) data, the inner error bar refers to the LSB catalog only while the outer error bar is due to the uncertainty in the space density of $\mu_0^*$ disks. Line: the surface brightness distribution suggested by Freeman (1970). This underpredicts the number of galaxies with $\mu_0 > 23$ mag arcsec$^{-2}$ by over 5 orders of magnitude.